\documentclass{article}

\usepackage{amsmath}
\usepackage{amssymb}
\usepackage{amsthm}
\usepackage{graphicx}
\usepackage{multirow}
\usepackage{color}
\usepackage{cite}

\title{Real-value and confidence prediction of protein backbone dihedral angles through a hybrid method of clustering and deep learning}
\author{Yujuan Gao$^{1,2}$, Sheng Wang$^{2}$, Minghua Deng$^{1,3,4*}$, Jinbo Xu$^{2}\footnote{Corresponding authors}$}

\begin{document}
\maketitle
\bibliographystyle{unsrt} 
\noindent$^1$ Center for Quantitative Biology, Peking University, Beijing, China\\
$^2$ Toyota Technological Institute at Chicago, Chicago, Illinois, United States of America\\
$^3$ School of Mathematical Sciences, Peking University, Beijing, China\\
$^4$ Center for Statistical Sciences, Peking University, Beijing, China\\

\begin{abstract}

\emph{Background.}
  Protein dihedral angles provide a detailed description of protein local conformation. Predicted dihedral angles can be used to narrow down the conformational space of the whole polypeptide chain significantly, thus aiding protein tertiary structure prediction. However, direct angle prediction from sequence alone is challenging.

\emph{Method.}
 In this study, we present a novel method to predict real-valued angles by combining clustering and deep learning. That is, we first generate certain clusters of angles (each assigned a label) and then apply a deep residual neural network to predict the label posterior probability. Finally, we output real-valued prediction by a mixture of the clusters with their predicted probabilities. At the same time, we also estimate the bound of the prediction errors at each residue from the predicted label probabilities.

\emph{Result.}
  In this article, we present a novel method (named RaptorX-Angle) to predict real-valued angles by combining clustering and deep learning.
  Tested on a subset of PDB25 and the targets in the latest two Critical Assessment of protein Structure Prediction (CASP), our method outperforms the existing state-of-art method SPIDER2 in terms of Pearson Correlation Coefficient (PCC) and Mean Absolute Error (MAE).
  Our result also shows approximately linear relationship between the real prediction errors and our estimated bounds. That is, the real prediction error can be well approximated by our estimated bounds.

  \emph{Conclusions.}
  Our study provides an alternative and more accurate prediction of dihedral angles, which may facilitate protein structure prediction and functional study.

\end{abstract}
\section{Introduction}
It has been shown that sequences contain rich information for protein tertiary structure prediction as well as functional study \cite{marks2012protein, de2013emerging}. But it is challenging to directly predict tertiary structure from primary sequence, so the hierarchical approach has been widely accepted as one of the most efficient methods. That means to transform the ultimate goal into several sub-problems, such as secondary structure prediction, solvent accessibility prediction, residue-residue contact prediction, etc. \cite{kurgan2011structural} reviewed the progress in the field of intermediate state or one-dimensional property prediction. It has been shown that predicted secondary structure is useful in the prediction of disordered and flexible regions, fold recognition and function prediction. However, secondary structure states are described as discrete classes and there is no clear boundary between coil and helical/strand states.
    It is a significant step towards establishing the structure and function of a protein to predict local conformation of the polypeptide chain. The local structural bias information restricts the possible conformations of a sequence segment and therefore narrows down the conformation space of the whole polypeptide chain significantly. Thus, prediction of dihedral angles is especially useful for protein tertiary structure prediction.

    On the whole, dihedral angle prediction may benefit protein tertiary structure prediction in several aspects.
    Firstly, dihedral angle prediction may act as substitute or supplement for secondary structure prediction \cite{wood2005protein, kountouris2010predicting, faraggi2012spine}. Secondly, It can be used in generation of sequence/structure alignment. For one thing, it can be directly applied to structure alignment methods based on dihedral angles \cite{miao2008tali, jung2011validity} and may aid refinement of target-template structure alignment. For another, considering predicted angles to refine multiple sequence alignment may narrow the gap between sequence and structure alignment, thus aiding de novo prediction of structural properties. In addition, dihedral angle prediction may also find applications in protein structure prediction that includes but not limits to fold recognition approaches \cite{zhang2002fold, zhang2008sp}, fragment-free tertiary structure prediction \cite{faraggi2009predicting}, tertiary structure refinement and structure quality assessment \cite{sims2006method} and functional study, such as ligand-binding site prediction \cite{cao2016improving}.

    There are mainly two kinds of problems in dihedral angle prediction: angle region prediction and real value prediction, which corresponds to two different representations of protein backbone local structural bias.

    Initially, Ramachandran basin is an intuitive description of local structural bias \cite{ramachandran1968conformation}. A Ramachandran basin is a specific region of a Ramachandran plot and illustrates the preference of torsion angle values. Each angle pair can be assigned a basin label. With more basins, the assignment would be harder but the representation would be more accurate and vice versa.  Colubri \emph{et al.} tested the ability to recover the native structure from a given basin assignment for each residue to investigate the level of representation required to simulate folding and predict structure, resulting in five basins \cite{colubri2006minimalist}. Gong \emph{et al.} partitioned $\phi,\psi$-space into a uniform grid of 36 squares, each $60^{\circ}\times60^{\circ}$, thus resulting in 36 basins, and showed that they successfully reconstructed six proteins solely from their mesostate (basin label) sequences \cite{gong2005building}. There are also some other methods to define basins and do angle region prediction with different definitions of basins \cite{dowe1995circular, kuang2004protein, zimmermann2006support, zhang2013accurate}. Although it is vital to determine the proper number of regions and clearly define the boundary, a universal algorithm to generate Ramachandran basins and assign basin labels remains to be developed. In our study, k-means clustering serves as the basin generator and label assigner.

    While Ramachandran basin provides an overall description of conformation, it is a coarse-grained representation and lacks statistical explanations describing the torsion angle distributions of each basin. In consideration of the circular nature of angles, traditional parametric or non-parametric density estimation methods cannot work properly to approximate Ramachandran distributions. Fortunately, directional distributions such as von Mises distribution could solve the problem \cite{singh2002probabilistic}. Bivariate von Mises distribution (mixtures) has been used to model protein dihedral angle distribution \cite{mardia2007protein, li2008fragment}, which removes arbitrariness in defining the boundary between discrete states. In this study, we assume angle pairs in each basin follow a bivariate von Mises distribution to derive the log-likelihood of each clustering.

    Thanks to the rapid growth of Protein Data Bank and computational and algorithmic development in machine learning (especially deep learning), several supervised machine learning methods have been proposed to predict real values of dihedral angles. As $\phi$ values in $\alpha$-helices and $\beta$-sheets are quite similar, $\psi$ seems more informative. Wood \emph{et al.} first developed a method DESTRUCT for prediction of real-valued dihedral angle $\psi$ and used this information for prediction of the protein secondary structure with high accuracy \cite{wood2005protein}. Wu \emph{et al.}proposed a composite machine-learning algorithm called ANGLOR to predict real-value protein backbone torsion angles from protein sequences \cite{wu2008anglor}. The input features of ANGLOR include sequence profiles, predicted secondary structure and solvent accessibility. The mean absolute error (MAE) of the $\phi$/$\psi$ prediction was reported to be $28^{\circ}$/$46^{\circ}$. Later Song \emph{et al.} developed TANGLE based on a two-level support vector regression approach using a variety of features derived from amino acid sequences, including the evolutionary profiles and natively disordered region as well as other global sequence features \cite{song2012tangle}. The MAE of the $\phi$/$\psi$ was $27.8^{\circ}$/$44.6^{\circ}$. Xue \emph{et al.} established a neural network method called Real-SPINE, with sequence profiles generated from multiple sequence alignment and predicted secondary structures as inputs \cite{xue2008real}. In 2015, they presented SPIDER2 \cite{heffernan2015improving} by improving SPIDER \cite{lyons2014predicting} through iterative learning, which used a deep artificial neural network (ANN) with three hidden layers of 150 nodes. They fed the predicted torsion angles of last layer as the input to the following generation and reported $19^{\circ}$ and $30^{\circ}$ for mean absolute errors of backbone $\phi$ and $\psi$ angles, respectively. As it is impossible to introduce all methods here, interested readers can refer to excellent reviews \cite{singh2014evaluation, zimmermann2017backbone}.

    Although there has been tremendous development, their performance is still limited by their shallow architectures. Inspired by the excellent performance of convolution neural network in predicting secondary structure \cite{wang2016protein} and order/disorder regions \cite{wang2015deepcnf} and also the success of residual framework to do contact prediction \cite{wang2017accurate}, we adopt the ultra deep residual framework of convolutional neural network to do k-means basin label probability prediction.

    However, even though a protein backbone conformation can be highly accurately rebuilt from its respective native dihedral angles, accumulation of errors in predicted angles can lead to large deviation in three-dimensional structures, which prevents angle prediction from its direct use in building protein structures \cite{heffernan2015improving}. It is of great significance to produce the corresponding confidence scores for the real value predictions, i.e., we need to know the confidence level of the predictions. Otherwise the effect of predicted dihedral angles as restraints for three dimensional structure prediction would be limited \cite{faraggi2009improving}. Zhou \emph{et al.} had developed SPIDER2 \cite{heffernan2015improving} to predict real-valued angles and then separately SPIDER2-Delta \cite{gao2016predicting} to predict error of those predicted structural properties. Here we describe a simple hybrid technique to predict angles and confidence scores simultaneously.

    Another problem that need to be considered is the periodicity of angles. For example, if an angle $\theta=179^{\circ}$ is predicted to be $-179^{\circ}$, the error would be treated as $358^{\circ}$ instead of $2^{\circ}.$ There are some approaches proposed to reduce the impact of cyclic nature of angles. One was angle shifting to reduce confusion at $0^{\circ}$ and $360^{\circ}$ (or $-180^{\circ}$ and $180^{\circ}$), e.g., shifting $\psi$ by $100^{\circ}$ and $\phi$ by $-10^{\circ}$ \cite{xue2008real} or adding $100^{\circ}$ to the angles between $-100^{\circ}$ and $180^{\circ}$ and adding $460^{\circ}$ to the angles between $-180^{\circ}$ and $-100^{\circ}$ \cite{faraggi2009improving}. But the improvement was limited and strongly depended on the angle range. For amino acids such as alanine that had minimal residues in the affected range, angle shifting made little difference \cite{singh2014evaluation}. A better choice was to take advantage of the inherent angle periodicity of trigonometric functions, that is, mapping the angles to their sine and cosine values \cite{heffernan2015improving}, which has achieved best performance so far. Inspired by this, we deal with equivalent trigonometric representations of dihedral angle pairs, rather than real value angles.

    Considering dihedral angles share similar patterns in alpha helix and beta strand, the acceptable $(\phi, \psi)$ patterns are limited. Moreover, it is much easier to do classification than regression. Also indebted to mixture models and Expectation-maximization algorithm, we develop a hybrid method of k-means clustering and deep learning to do angle prediction, combining advantages of discrete and continuous representation of dihedral angles.
    Specifically, we firstly generate a set of clusters of $(\phi, \psi)$ from training data, in which we could get the distribution of each cluster; then we use deep learning methods to predict discrete labels; lastly we predict real value angles by mixing empirical clusters with their predicted probabilities. We employ a residual framework of convolutional neuron network in RaptorX-Angle to predict the cluster label probabilities.
    We test our method on filtered PDB25 dataset as well as CASP (Critical Assessment of protein Structure Prediction) targets and compare with other three state-of-art methods.
    Tested on the subset of PDB25, our method gains about $0.5^{\circ}$ and $1.4^{\circ}$ for $\phi$ and $\psi$ better MAE than SPIDER2, currently among the best backbone angle predictors. Our method also performs better than SPIDER2 on the CASP11 and CASP12 test targets. The advantage is even more obvious when looking into detailed secondary structural regions.

\section{Methods}

    \subsection{K-means clustering of angle vectors}
        \paragraph{Genearating k-means ``centers'' from angle vectors.}
        For a dihedral angle pair $(\phi, \psi)$, we can equivalently denote it by an angle vector
        $$\mathbf{v}=(\cos(\phi), \sin(\phi), \cos(\psi), \sin(\psi)).$$
        Conversely, given the vector representation $\mathbf{v}$, we can easily derive the corresponding angles $\phi$ and $\psi$ (Supplementary Material S1.1).
        We run k-means on angle vectors to cluster dihedral angle pairs in training set into $K=10, 20, \ldots, 100$ clusters. Then we normalize the $K$ centres $\{\mathbf{C}_k\}_{k=1}^K$ and get the final ``centers'' $\{\widetilde{\mathbf{C}}_k
        =(\widetilde{c}_{k0}, \widetilde{c}_{k1}, \widetilde{c}_{k2}, \widetilde{c}_{k3})\}_{k=1}^K$, so that each ``centre'' $\widetilde{\mathbf{C}}_k$ is a valid representation for some angle pair (Supplementary Material S1.2).

        \paragraph{Predicting ``true'' labels from k-means.}
        Given the $K$ normalised vector ``centres'' $\{\widetilde{\mathbf{C}}_k\}_{k=1}^{K}$, we could assign the ``true'' label for each dihedral angle pair as the one whose corresponding normalised centre was closest to its respective vector representation. Then the ``true'' labels can be used as the training labels to build a deep learning model as a classifier to predict labels for testing data.
    \subsection{Deep learning model details}

        \paragraph{Deep Convolutional Neural Network (DCNN).} DCNN consists of multiple convolutional blocks.
            A convolutional block is a neural network that implements a composite of linear convolution and nonlinear activation transformation. Convolution is used in place of general matrix multiplication, which can better capture local dependency. It has been widely accepted that protein torsion angles strongly depend on neighbour residues \cite{betancourt2004local, keskin2004relationships, jha2005helix}. So DCNN is ideal to abstract angle information from sequence.

\begin{figure}
          \centering
          \includegraphics[scale=0.45]{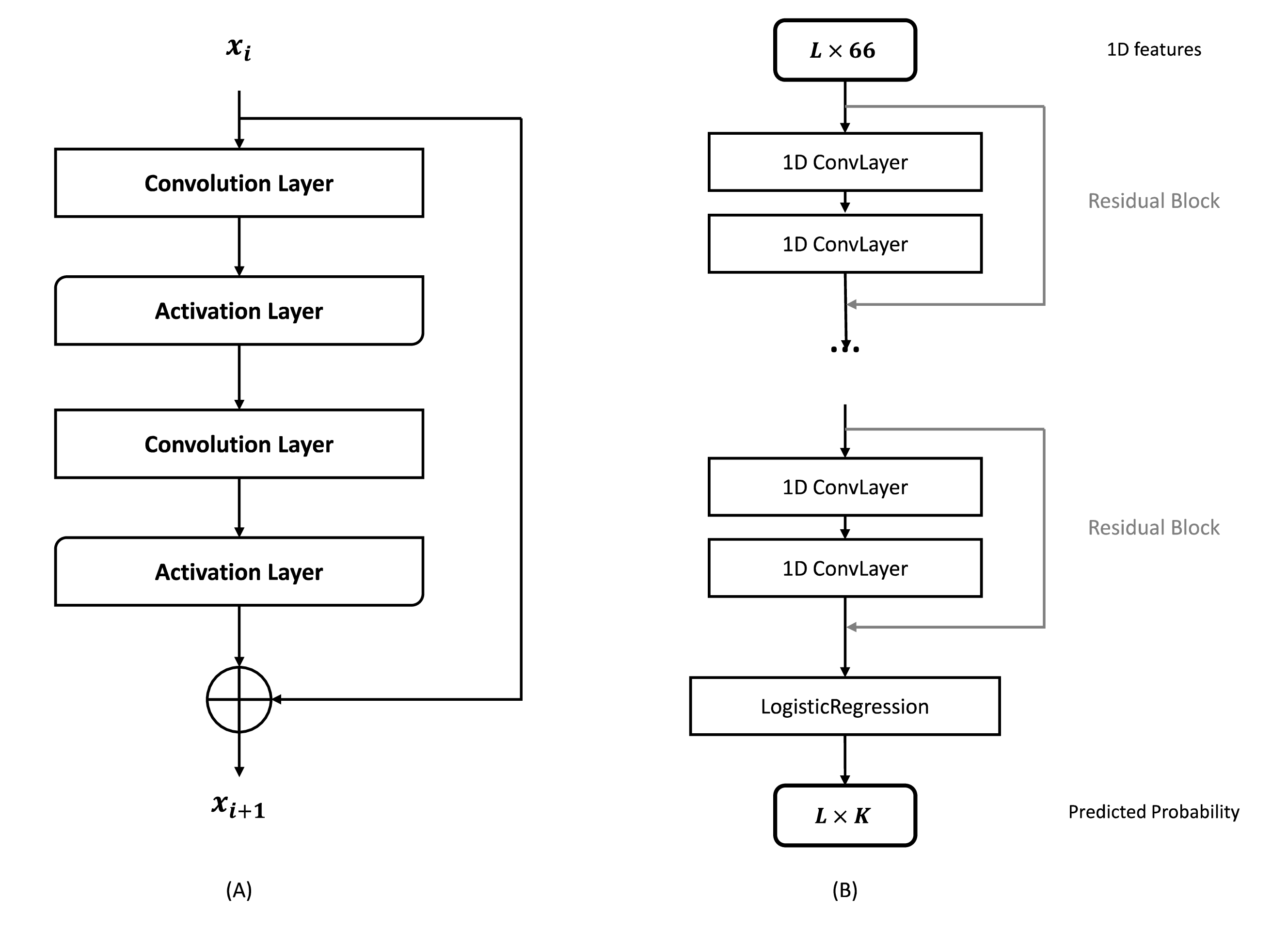}
          \caption{Illustration of the ResNet model in RaptorX-Angle. (A) A building block of ResNet with $x_i$ and $x_{i+1}$ being input and output, respectively. (B) The ResNet model architecture as a classifier with stacked residual blocks and a logistic regression layer. Here $L$ is the sequence length of the protein or total number of residues under prediction and $K$ is the number of clusters.}
          \label{illustration}
        \end{figure}

        \paragraph{Residual Network (ResNet).} DCNN can integrate features in hierarchical levels and some work has shown the significance of depth \cite{szegedy2015going}. However, with the depth increasing, accuracy gets saturated and even degraded. That is because adding more layers may lead to higher training error as identity mapping is difficult to fit with a stack of nonlinear layers \cite{srivastava2015training}. ResNet was proposed as a residual learning framework to ease the training of substantially deeper networks \cite{he2016deep}.
        Figure \ref{illustration} demonstrates the basic architecture of ResNet in RaptorX-Angle. Figure \ref{illustration}(A) is a residual block, which consists of 2 convolution layers and 2 activation layers, and the ResNet consists of stacked residual blocks (Figure \ref{illustration}B). The activation layer conducts a simple nonlinear transformation of its input depending on the activation function with no additional parameters. In this work, we used the ReLU activation function \cite{nair2010rectified}.

        \paragraph{Logistic Regression layer.} DCNN and ResNet can capture information from data and output abstract features. To do classification for residues, a logistic regression layer is added as the final layer in RaptorX-Angle, which could output the marginal probability of $K$ labels (Figure \ref{illustration}(B)).

        \paragraph{Loss function.} We train model parameters through maximizing the probability of angle pairs belong to the ``true'' labels. Naturally, the loss function is defined as the negative log-likelihood averaged over all residues of the training proteins.

        \paragraph{Regularization and optimization.}
        As is widely used in machine learning, the log-likelihood objective function is penalized with a $L_2$-norm of the model parameters to prevent overfitting. Thus, the final objective function has two items: loss function and regularization item, with a regularization factor $\lambda$ to balance the two items. That is, the final objective function is:
        $$ \max_{\theta}\quad\log{P_{\theta}(Y|X)}-\lambda\|\theta\|^2$$
        where $X$ is the input features, $Y$ is the output labels, $\theta$ is the model parameters and $\lambda$ is the regularization factor used to balance the log likelihood and regularization.
        We use Adam \cite{kingma2014adam} to minimize the objective function, which usually can converge within 20 epochs. The whole algorithm has been implemented by Theano \cite{bergstra2010theano} and mainly run on a GPU card.

        \paragraph{Input features.}
        For each residue in each protein sequence, we generate a total of 66 input features,
        of which 20 from position specific scoring matrix(PSSM) of PSI-BLAST \cite{camacho2009blast+}, 20 from position-specific frequency matrix (PSFM) of HHpred \cite{soding2004protein, remmert2012hhblits},
        20 from primary sequence, 3 from predicted solvent accessibility (ACC) and 3 from predicted secondary structure(SS) probabilities (Supplementary Material S1.3).

    \subsection{Predicting real-value angles from predicted marginal probability}
    From the last logistic regression layer of the deep learning model, we could predict the marginal probability $\mathbf{P}=(p_1,p_2,\ldots,p_{K})$ of an angle pair for each label. We use the marginal probability rather than the single predicted label to reduce bias. Concretely, we calculate the weighted mean by:
    $$\widehat{\mathbf{v}}=(v_0,v_1,v_2,v_3)=\sum_{k=1}^{K}p_k\widetilde{\mathbf{C_k}},$$
    Finally, we normalise $\widehat{\mathbf{v}}$ to get
    $$\widehat{\cos(\phi)}=\frac{v_0}{\sqrt{ {v_0}^2 + {v_1}^2 }}, \widehat{\sin(\phi)}=\frac{v_1}{\sqrt{ {v_0}^2 + {v_1}^2 }},$$
    $$\widehat{\cos(\psi)}=\frac{v_2}{\sqrt{ {v_2}^2 + {v_3}^2 }}, \widehat{\sin(\psi)}=\frac{v_3}{\sqrt{ {v_2}^2 + {v_3}^2 }}.$$
    and we could derive the predicted real values $\widehat{\phi},\widehat{\psi}$ from this angle vector (Supplementary Material S1.1).
    We also tried to predict real-value angles from labels with top $R$($R<K$) probabilities when $K$ is well chosen (Supplementary Material S2.3).

    \subsection{Programs to compare and evaluation metrics}
    We compare our method with three available standalone softwares SPIDER2 \cite{heffernan2015improving}, SPINE X \cite{faraggi2009predicting}, and ANGLOR \cite{wu2008anglor}. All the programs are run with parameters suggested in their respective papers.

    We evaluate the performance by Pearson Correlation Coefficient (PCC) and Mean Absolute Error (MAE) as described by \cite{kountouris2009prediction}, for assessing the prediction of $\phi$/$\psi$ angles. Considering the periodicity of angles, PCC is calculated between the cosine (sine) values of predicted and experimentally determined angles. MAE is the average absolute difference between predicted and experimentally determined angles. The periodicity of an angle has been taken care of by utilizing the smaller value of the absolute difference $d(=|\theta_{pred}-\theta_{exp}|)$ and $360-d$ for average, where $\theta_{pred}$ is the predicted angle and $\theta_{exp}$ is the true angle value.

\section{Results}
    \subsection{Datasets}
    We use the targets from PDB25 updated in February, 2016. The set consists of 10820 non-redundant protein chains, in which any two chains share no more than 25\% sequence identity. To remove impact of disordered regions, we filter out proteins with internal disordered regions by DSSP \cite{kabsch1983dictionary}. Finally we get 7604 proteins. We then randomly select 5070 proteins as the candidate training set, 1267 as validation set (VL1267, see VL1267\_list.txt) and the remaining 1267 as test set (TS1267, see TS1267\_.txt).
    We also test on 85 CASP11 targets (see casp11\_list.txt) and the latest 40 CASP12 targets (see casp12\_list.txt) with publicly released native structures.
    To remove redundancy between training proteins and CASP targets, we run MMseqs2 \cite{steineggersensitive}, which is similar but more sensitive and faster than BLAST (PSI-BLAST) for protein sequence homology search, with $seqID$ cutoff 0.25 and also E-value cutoff 0.001 to filter 5070 the candidate training proteins, resulting in 5046 training proteins (TR5046, see TR5046\_list.txt).

    \subsection{Choosing a proper number of clusters}
    A vital problem is how to select the number of clusters, which can be reduced to defining measures for clustering evaluation. Here we adopt two measures: (i) entropy loss based on discrete distribution; (ii) loglikelihood based on continuous distribution to evaluate 10 different clusterings ($K=10,20,\ldots,100$). Firstly, we do k-means clustering on TR5046 and get $K$ empirical clusters. Secondly, we train the deep learning models and do classification on VL1267, then we can obtain the predicted marginal probability of the $K$ clusters $\mathbf{P_i}=(p_{i1},p_{i2},\ldots,p_{iK}),i=1,2,\ldots,N$, where $i$ is the index of residue and $N$ is the total number of residues in VL1267.

    \paragraph*{Entropy loss.} Entropy $H(\cdot)$ is always used to measure the information of a distribution. From k-means clustering on TR5046, the background distribution among clusters $\mathbf{P_0}=(p_{01},p_{02},\ldots,p_{0K})$ could be derived. Then the entropy loss of this clustering on VL1267 can be calculated as the mean difference between entropy of background distribution and predicted marginal distribution:
    \begin{eqnarray*}
    EL & = & \frac{1}{N}\sum_{i=1}^N(H(\mathbf{P_0})-H(\mathbf{P_i}))\\
       & = & \frac{1}{N}\sum_{i=1}^N(\sum_{k=1}^Kp_{0k}\log(p_{0k})-\sum_{k=1}^Kp_{ik}\log(p_{ik}))
    \end{eqnarray*}
    which can roughly evaluate the information gain from the clustering. Here $N$ is the number of residues in VL1267.

    \paragraph{Loglikelihood.} To demonstrate the detailed information of each cluster, we need a continuous angular(circular) distribution defined on the torus. Mixture bivariate von Mises distributions are successfully used to describe the local bias of torsion angle pair $(\phi, \psi)$ \cite{singh2002probabilistic, mardia2007protein, li2008fragment}, we assume that angle pairs belong to the same cluster $k$ obey a common bivariate von Mises distribution $f_k$ with parameters $\Theta_k=(\kappa_1^k, \kappa_2^k, \kappa_3^k, \mu^k, \nu^k)$. Here,

    \begin{eqnarray*}
    f_k(\phi, \psi) & = & c(\kappa_1^k, \kappa_2^k, \kappa_3^k)\exp\{\kappa_1^k\cos(\phi-\mu^k)\\
       &  &  + \kappa_2^k\cos(\psi-\nu^k) + \kappa_3^k\cos(\phi-\mu^k-\psi+\nu^k)\}
    \end{eqnarray*}
    where $\mu^k$ and $\nu^k$ are the mean value of $\phi$ and $\psi$, respectively; $\kappa_1^k, \kappa_2^k$ are the concentrations, $\kappa_3^k$ allows for the dependency between the two angles and $c(\kappa_1^k, \kappa_2^k, \kappa_3^k)$ is a normalization constant:
    \begin{eqnarray*}
    c(\kappa_1^k, \kappa_2^k, \kappa_3^k)  =  (2\pi)^2\big\{ I_0(\kappa_1^k)I_0(\kappa_2^k)I_0(\kappa_3^k)  + 2\sum_{p=1}^{\infty}I_p(\kappa_1^k)I_p(\kappa_2^k)I_p(\kappa_3^k) \big\}
    \end{eqnarray*}
    in which $I_p(\kappa)$ is the modified Bessel function of the first kind and order $p$.
    Parameters $\{\Theta_k=(\kappa_1^k, \kappa_2^k, \kappa_3^k, \mu^k, \nu^k)\}_{k=1}^K$ can be intuitively estimated from the empirical clusters $\{(\phi, \psi)_k\}_{k=1}^K$ \cite{hamelryck2012bayesian}. Then the density function for the torsion angle pair $(\phi, \psi)$ can be approximately described as:
    $$f(\phi, \psi)=\sum_{k=1}^K p_kf_k(\phi,\psi)$$
    where $p_k$ is the predicted marginal probability of $(\phi, \psi)$ belongs to cluster $k$.
    Then the log-likelihhod for the VL1267 can be calculated as:
    $$LL=\frac{1}{N}\sum_{i=1}^N\log{f(\phi_i, \psi_i)}=\frac{1}{N}\sum_{i=1}^N\log{\sum_{k=1}^K p_{ik}f_k(\phi_i,\psi_i)}$$

    \paragraph{Selecting proper $K$.} Figure 2 shows the result of entropy loss and loglikelihood with respect to the number of clusters. As expected, the loglikelihood increases along with $K$, which means it can better describe the data with more clusters. But when $K$ goes larger than 30, there is an obvious decrease in entropy loss. Maybe that is because the more clusters are used, the more challenging it would be to do angle prediction. As there is a soaring information gain when $K$ goes from 10 to 20 and little difference when K increases from 20 to 30,
    we test every single clustering between 20 and 30 and there is no significant benefit with more clusters. So we just choose $K=20$ to do following studies.
    \begin{figure}[!tpb]
      \centerline{\includegraphics[scale=0.55]{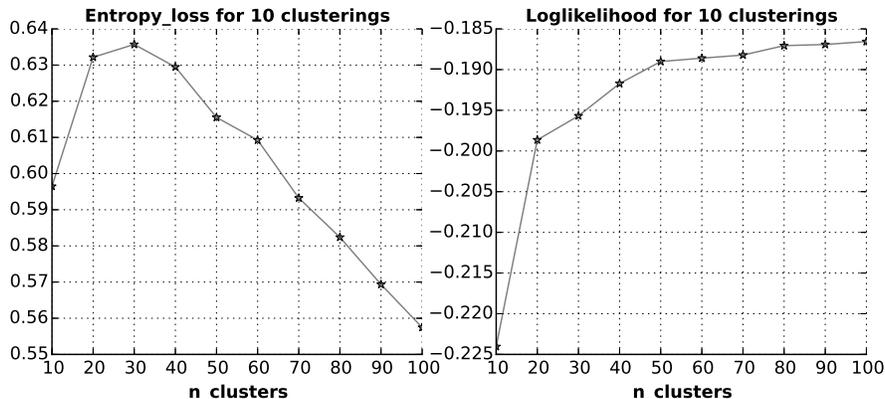}}
      \caption{Selecting proper number of clusters. Left: entropy loss of discrete label probabilities; Right: loglikelihood of mixture bivariate von Mises distribution.}\label{fig:sel}
    \end{figure}

    \subsection{Feature contribution study}
    The features can be divided into three categories: sequence information including amino acid (aa) and profile, predicted secondary structure (SS) and solvent accessibility (ACC). Sequence profile information are generated from PSI-BLAST (PSSM) and HHpred (PSFM) (See Supplementary Material S1.3 for more details). To test the impact of different feature combinations, we design six experiments: (1) basic1 = 20 PSSM + 20 aa; (2) basic2 = 20 PSFM + 20 aa; (3) basic = 20 PSSM + 20 PSFM + 20 aa; (4) basic + 3 ACC; (5) basic + 3 SS; (6) basic + 3 ACC + 3 SS.
    The network architecture is fixed as $N_{layers}=5, N_{nodes}=100, halfWinSize=3$ (ResNet 5-100-3), and the regularization factor is fixed to be 0.0001.

    Table \ref{Tab:contribution} shows the MAE performance of different feature combinations on TS1267. From the first three experiments with only sequence information involved, there is little performance difference between PSSM and PSFM, and the combination of PSSM and PSFM gains the best accuracy. So PSSM and PSFM are complementary and both unignorable. ACC and SS both contribute significantly and also the combination gain the best accuracy. Finally we use the whole set of features.
\begin{table}[h!]
  \caption{The Mean Absolute Error of different feature combinations with ResNet 5-100-3 on TS1267.}
  \label{Tab:contribution}
  {
  \begin{tabular}{ccccccccc}\hline
    ID & Phi & Psi &  Phi\_H & Psi\_H & Phi\_E & Psi\_E & Phi\_C & Psi\_C \\ \hline
    (1) & 19.97 & 31.97 & 9.82 & 17.57 & 20.70 & 26.97 & 29.84 & 49.66 \\
    (2) & 20.02 & 31.78 & 9.86 & 17.68 & 20.46 & 26.38 & 30.10 & 49.39 \\
    (3) & 19.27 & 30.04 & 9.11 & 15.58 & 19.70 & 24.64 & 29.35 & 48.02 \\
    (4) & 19.08 & 29.30 & 9.07 & 15.44 & 19.36 & 23.18 & 29.11 & 47.10 \\
    (5) & 19.19 & 28.73 & 8.56 & 13.76 & 19.29 & 22.43 & 31.00 & 47.95 \\
    (6) & \textbf{18.58} & \textbf{27.98} & \textbf{8.45} & \textbf{13.37} & \textbf{19.03} & \textbf{22.14} & \textbf{28.61} & \textbf{46.21} \\
    \hline

  \end{tabular}
  {\\ Phi and Psi denote MAE for all residues; \\
    Phi\_H and Psi\_H denote MAE for residues in helix region; \\
    Phi\_E and Psi\_E denote MAE for residues in beta strand region; \\
    Phi\_C and Psi\_C denote MAE for residues in coil region.\\
    $(\cdot)$ indicates the id of the feature contribution experiments.}
  }
\end{table}

    \subsection{Overall PCC performance of cosine values compared with other methods}
    To tune proper regularization factor and also network architectures, we perform 5-fold cross validation on TR5046 (Supplementary Material S2.1, S2.2). Finally, we choose an ensemble of 6 networks (Supplementary Material S2.2).
    We test our method on TS1267 and also the popular CASP targets, including 85 CASP11 targets and 40 CASP12 targets.
    Table \ref{Tab:pcc} shows the PCC performance of cosine values on the three benchmarks. RaptorX-Angle has gained the highest PCC on all datasets. We also evaluate PCC performance of sine values (See Supplementary Material S2.4) and get similar results.
\begin{table}[h!]
  \caption{Pearson correlation coefficient of cosine values between predicted and true angles.}
  \label{Tab:pcc}
  {
  \begin{tabular}{c|cc|cc|cc}
  \hline
  \quad & \textbf{TS1267} & \textbf{CASP11} & \textbf{CASP12} \\
  \quad &  $\cos(\phi)/\cos(\psi)$ & $\cos(\phi)/\cos(\psi)$ & $\cos(\phi)/\cos(\psi)$ \\ \hline
  RaptorX-Angle & \underline{0.7111}/\underline{0.7576} & \underline{0.6585}/\underline{0.7103} & \underline{0.6539}/\underline{0.6979} \\
  SPIDER2 & 0.6893/0.7427 & 0.6485/0.7095 & 0.6299/0.6761 \\
  SPINE X & 0.6410/0.6543 & 0.5015/0.4891 & 0.4990/0.5039 \\
  ANGLOR & 0.4775/0.6226 & 0.4437/0.5868 & 0.4431/0.5772 \\
  \hline
  \end{tabular}
  }
\end{table}

    \subsection{Overall MAE performance compared with other methods}
    Table \ref{Tab:MAE} shows the MAE performance on the three benchmarks in different secondary structural regions of our RaptorX-Angle comparing with other three methods. All methods have larger MAE on CASP targets than on TS1267. It is reasonable since CASP targets are usually hard to predict. It can be seen that RaptorX-Angle performs the best on all benchmarks, with about $0.5^\circ$ and $1.4^\circ$ for $\phi$ and $\psi$ better MAE on both TS1267 and CASP12 and slightly better performance on CASP11 than the second best method SPIDER2.
    We perform Student's t test of absolute errors between RaptorX-Angle and SPIDER2. As a result, the p-values for $\phi/\psi$ are $8.65e-12/2.79e-33$, $5.13e-2/8.36e-2$ and $1.28e-5/2.59e-8$ on TS1267, CASP11 and CASP12, respectively.
    That is , the advantage of RaptorX-Angle over SPIDER2 on TS1267 and CASP12 is statistically more significant than on CASP11.
    These results demonstrate the rationality of representing the Ramachandran plot with a limited number of clusters, say 20 clusters, and also reflect the power of deep learning methods.
\begin{table}[h!]
    \caption{Mean absolute error of four methods for different secondary structural regions on three benchmarks: TS1267, 85 CASP11 targets and 40 CASP12 targets.}
    \label{Tab:MAE}
    {
    \begin{tabular}{ccccccccc}
    \hline
    ($^{\circ}$) & Phi & Psi &  Phi\_H & Psi\_H & Phi\_E & Psi\_E & Phi\_C & Psi\_C \\ \hline
    \textbf{TS1267} & & & & & & & &\\
    RaptorX-Angle &  $\underline{18.08}$ & $\underline{26.68}$ & $\underline{8.35}$ & $\underline{12.98}$ & $\underline{18.24}$ & $\underline{20.94}$ & $\underline{27.88}$ & $\underline{44.11}$\\
    SPIDER2 &18.57 &28.02 &8.59 &14.52 &19.28 &23.09 &28.28 &44.73\\
    SPINE X & 20.31 &34.05 &9.32	&16.69 &22.23 &31.23 &30.32 &53.42\\
    ANGLOR & 24.01 &43.59 &9.29 &26.41 &27.47 &40.88 &36.89 &62.72\\
    \\
    \textbf{CASP11} & & & & & & & &\\
    RaptorX-Angle &  $\underline{20.00}$ & $\underline{30.14}$ & \underline{9.49} & $\underline{15.65}$ & $\underline{18.82}$ & $\underline{23.58}$ & \underline{29.87} & 46.89\\
    SPIDER2 & 20.18 & $30.32$ & 9.53 & 16.05 & 19.77 & 24.50 &29.88 & \underline{46.84} \\
    SPINE X  & 24.85 & 46.58 &13.57 &29.65 &26.25 &43.65 &33.88 &63.49 \\
    ANGLOR  &25.69 &46.17 &9.99 &27.72 &28.08 &43.85 &37.96 &64.03 \\

    \\
    \textbf{CASP12} & & & & & & & &\\
    RaptorX-Angle & $\underline{20.69}$ & $\underline{32.73}$ &9.28 & $\underline{16.73}$ & $\underline{19.94}$ & $\underline{26.06}$ & $\underline{31.22}$ & $\underline{51.02}$ \\
    SPIDER2 & 21.13 & 34.17 & $\underline{9.13}$ & 17.19 & 21.35 & 28.56 & 31.95 & 52.76 \\
    SPINE X  & 24.85 & 46.57 & 11.52 & 26.34 & 26.98 & 46.04 & 35.85 & 65.33 \\
    ANGLOR  & 25.79 & 47.37 & 9.69 & 28.81 & 29.11 & 44.79 & 38.65 & 65.74 \\
    \hline
    \end{tabular}
    }{\\ Same notations with Table \ref{Tab:contribution}}

\end{table}

    \subsection{Mean absolute error performance study in VL1267}
\begin{figure}[!tpb]
          \centerline{\includegraphics[scale=0.55]{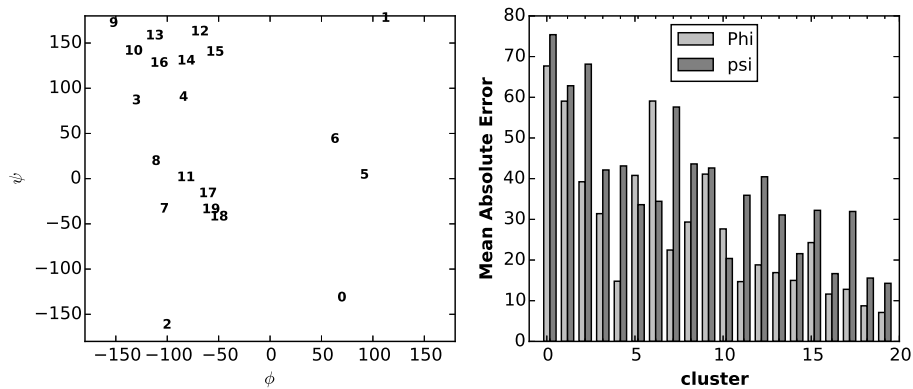}}
          \caption{Mean absolute error performance for different clusters in VL1267.
                   Left: visualization of 20 cluster centers on the Ramachandran plot with smaller number indicating smaller size.
                   Right: mean absolute error for different clusters.
                   }
          \label{fig:MAE_cluster}
        \end{figure}

    In methodology, the conversion from angle pair to trigonometric vector is nonlinear, the prediction error may depend on angles. And in biology, prediction error may differ for different amino acids with different microscopic biochemical properties, and also for different protein classes with different macroscopic structures. So we perform detailed studies on prediction error in VL1267.
        \paragraph{Study Mean absolute error performance for different clusters.}
        As each cluster corresponds to a certain angle region, we calculate the MAE for each cluster in VL1267. We observe that the 20 clusters are well consistent with Ramachandran plot and also the two peaks for $\phi$ and $\psi$ \cite{faraggi2009predicting} (Figure~\ref{fig:MAE_cluster} Left). And the prediction errors differ a lot between clusters. It turns out that clusters with more residues in coil region tend to result in larger prediction errors. Moreover, prediction error for $\phi$ is smaller than for $\psi$. But there are three uncommon clusters with larger MAE for $\phi$, i.e., 5, 6 and 10 (Figure~\ref{fig:MAE_cluster} Right). Clusters 5 and 6 are totally in one of the peak areas in Ramachandran plot, which may indicate some interesting biological discoveries.

        \paragraph{Mean absolute error performance for each amino acid type.}
        As different amino acids have different stereochemical and physiochemical properties, they are anticipated to have different degrees of difficulty for the torsion angle prediction. In Table~\ref{Tab:MAE_aa}, we examine the MAE performance for each of 20 amino acid types.
        Glycine, with no side-chain atom except for a proton, has least steric restriction to backbone dihedral angle motions. As a result, it has the largest prediction error ($43.32^\circ/39.59^\circ$ for $\phi/\psi$).
        In contrast, Proline has the least MAE $(8.84^\circ)$ for $\phi$ but has an unusually large MAE $(33.00^\circ)$ for $\psi$ prediction due to its special side-chain structure, which is consistent with \cite{wu2008anglor}.     In addition, three of the amino acids (Ile, Leu and Val) with the smallest MAE are all hydrophobic.
\begin{table}[h!]
  \caption{Mean absolute error performance for each amino acid type in VL1267.}
  \label{Tab:MAE_aa}
  {
  \begin{tabular}{ccccc}
  \hline
    Amino acids & Abundance & Frequency$(\%)$ & $\phi(^\circ)$ & $\psi(^\circ)$ \\ \hline
    A (Ala) & 22527 & 8.46 & 13.87 & 22.92 \\
    C (Cys) & 3151 & 1.18 & 20.50 & 28.66 \\
    D (Asp) & 15946 & 5.99 & 20.71 & 30.80\\
    E (Glu) & 18326 & 6.89 & 14.75 & 23.97\\
    F (Phe) & 10812 & 4.06 & 18.13 & 26.10\\
    G (Gly) & 19133 & 7.19 & \textbf{43.32} & \textbf{39.59}\\
    H (His) & 5989 & 2.25 & 22.04 & 31.12\\
    I (Ile) & 15302 & 5.75 & 12.79 & 20.12\\
    K (Lys) & 15299 & 5.75 & 16.71 & 25.83\\
    L (Leu) & 24731 & 9.29 & 12.49 & 21.37\\
    M (Met) & 5833 & 2.19 & 16.71 & 24.86\\
    N (Asn) & 11383 & 4.28 & 27.38 & 32.04\\
    P (Pro) & 11977 & 4.50 & \textbf{8.84} & \textbf{33.00}\\
    Q (Gln) & 10163 & 3.82 & 15.96 & 24.72\\
    R (Arg) & 13529 & 5.08 & 16.81 & 25.45\\
    S (Ser) & 15991 & 6.01 & 20.83 & 33.92\\
    T (Thr) & 14309 & 5.38 & 17.12 & 30.92\\
    V (Val) & 18612 & 6.99 & 13.70 & 20.94\\
    W (Trp) & 3854 & 1.45 & 18.05 & 27.61\\
    Y (Tyr) & 9287 & 3.49 & 18.83 & 27.02\\\hline
    Total & 266154 & 100 & 18.32 & 27.15\\
  \hline
  \end{tabular}
  }
\end{table}

\begin{figure}[!tpb]
          \centerline{\includegraphics[scale=0.5]{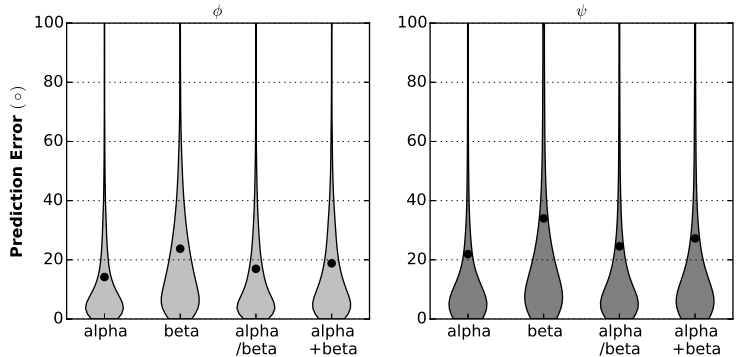}}
          \caption{Mean absolute error performance for different protein classes in VL1267. Left: for $\phi$ prediction. Right: for $\psi$ prediction.}
          \label{fig:MAE_class}
        \end{figure}

        \paragraph{Mean absolute error performance for different protein classes.}
        After studying on MAE performance in microcosmic view, we intend to study the performance for different macroscopical structures.
        We abstract 99, 117, 171, 117 proteins from VL1267 (resulted in 17696, 24874, 47304 and 19645 residues) in all $\alpha$, all $\beta$, $\alpha/\beta$ and $\alpha+\beta$ classes, respectively. We calculate the absolute error for every residue in each class. Figure~\ref{fig:MAE_class} shows the violin plot of prediction error for $\phi$ (Left) and $\psi$ (Right). A violin plot is similar to box plot except that it also shows the probability density of the data. We can see although the MAE for $\phi$ are smaller for all protein classes, prediction errors belong to each protein class have their own distribution pattern and the pattern is similar between $\phi$ and $\psi$. Overall, prediction errors are smallest in all $\alpha$ proteins and largest in all $\beta$ for both $\phi$ and $\psi$ predictions.

    \subsection{Estimating confidence score of predicted angles}
    Generally, variance $\sigma^2$ includes variance within cluster $\sigma_w^2$ and variance between cluster $\sigma_b^2$. To produce the confidence score of our predicted angles, we calculate the standard deviation from variances within a cluster. Specifically, for each cluster $k$, we can get the in-cluster variance $\sigma_k^2(\theta)$ from training data, where $\theta=\phi$ or $\psi$. Then we derive the variance of prediction by:
    $$var(\theta)=\sigma^2(\theta)=\sum_{k=1}^{K}p_k\sigma_k^2(\theta)$$

\begin{figure}[!tpb]
      \centerline{\includegraphics[scale=0.5]{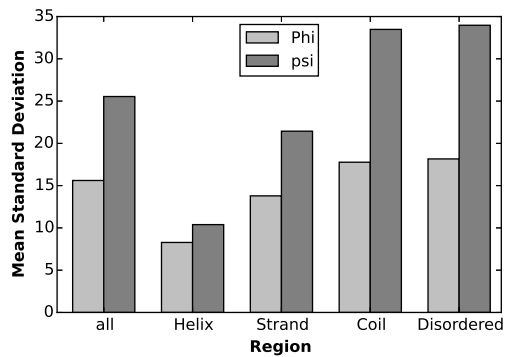}}
      \caption{Mean standard deviation for different secondary structural regions in TS1267.}
      \label{fig:std}
    \end{figure}

    Figure~\ref{fig:std} shows the mean standard deviation for $\phi$ and $\psi$ in different regions. As expected, the smallest variance appears in helix region, and then strand and lastly coil region. The standard deviation in disordered regions are rather large and quite similar to coil regions, which is consistent with our prior knowledge that disordered region resembles loop region and is rather flexible.

\begin{figure}[!tpb]
      \centerline{\includegraphics[scale=0.5]{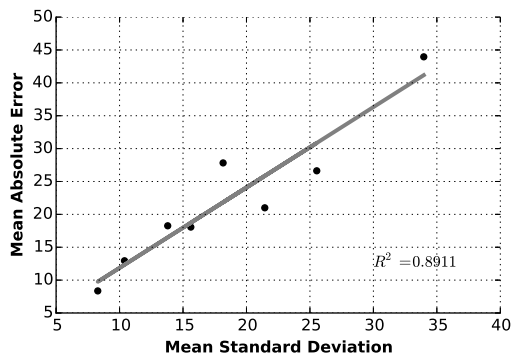}}
      \caption{Relationship between prediction error and standard deviation. Eight points are for two kinds of angles in four secondary structural regions (total, helix, strand, coil).} \label{fig:error_std}
    \end{figure}

    Figure~\ref{fig:error_std} demonstrates the relationship between MAE and mean standard deviation for $\phi$ and $\psi$ in different regions on VL1267. Roughly, the relationship is linear ($R^2=0.8911$). So the MAE can be bounded well by the standard deviation. We predict the error for each residue in each target from TS1267 and calculate corresponding Pearson and Spearman correlation coefficients ($PCC$ and $SCC$) between prediction errors and true errors, and also the mean absolute error for prediction errors (MAEPE). Finally, we obtain $PCC=0.3109, SSC=0.5427, MAEPE=13.94$ for $\phi$ and $PCC=0.2597, SCC=0.4751, MAEPE=26.21$ for $\psi$.
    We also try to fit two linear models for $\phi$ and $\psi$ separately on the all data points in VL1267 and get similar testing results. This indicates that the mean for different secondary structural regions almost contains enough information about the relationship between the estimated standard deviation and prediction error (Supplementary Material S2.6).

    \subsection{Computational cost analysis}
    All mentioned methods could do angle prediction target by target, so the computational cost is bounded by the longest protein (i.e., protein with the largest number of residues). To generate angle predictions for 1xdoA, the largest protein in TS1267 with 685 residues, it takes 726s, 123s, 370s and 524s for ANGLOR, spineX, SPIDER2 and RaptorX-Angle, respectively.

    As far as we see, the computational cost is mainly determined by method outline, network complexity, feature engineering and technical resources. ANGLOR is a composite method and the technology was not so developed at that time, it needs the most time. While spineX just adopted a simple network, SPIDER2 used more features iteratively in a more complex network and it takes longer than spineX.

    Compared with the second best SPIDER2, RaptorX-Angle used much deeper networks and also adopted profile information from hhblits (PSFM), besides PSSM from PSI-BLAST harnessed by spineX and SPIDER2. As a result, it takes SPIDER2 360s to generate features with 4 CPUs and 20s to predict angles using a CPU, while it takes RaptorX-Angle 385s to generate features with 4 CPUs, and 200s to predict angles from the features using a GPU card.

    However, we can integrate the features of a total batch of proteins and run them all at once. Actually, it just takes 750s to do angle prediction for all proteins in TS1267, while other methods needs many CPUs in parallel. Overall, our method is faster for prediction of many proteins and has gained better prediction accuracy.


\section{Discussion}

We have transformed the hard real-valued prediction problem into a discrete label assignment problem, which has simplified the problem and also gained better results. Overall, our RaptorX-Angle gains the best PCC in terms of cosine and sine of angles on all datasets. It has about $0.5^\circ$ and $1.4^\circ$ for $\phi$ and $\psi$ better MAE than the second best method SPIDER2 on a subset of PDB25. We have also calculated the two-state accuracy to see how much improvement there would be in large angle errors. RaptorX-Angle performs the best and has about 0.15 and 1 percent improvement over SPIDER2 for $\phi$ and $\psi$ on TS1267(See Supplementary Material S2.5). Our method also works very well on the CASP targets.
Moreover, we have estimated the prediction errors at each residue by a mixture of the clusters with their predicted probabilities. It has been shown that there is approximately linear relationship between the real prediction error and in-cluster standard deviation. That is a unique feature of our method.
In addition, we check the prediction for disordered regions. As there is no angle information, we just analyze the standard deviation and get quite large values and similar patterns to coil region. It is consistent with our prior knowledge that disordered region is rather flexible and resembles loop region. We also do comprehensive studies on prediction performance in VL1267, both in microscopic and macroscopic view.

This simple technique has gained better performance than other state-of-art methods. It demonstrates that for protein structures, the 20 clusters contain enough information for $(\phi, \psi),$ which is an efficient compression of information. The idea that to predict dihedral angles from clustering has turned out to be successful due to three aspects. The first is the continuous growth of the solved structures \cite{berman2000protein}, so we have enough training data. The second is the novel idea to predict real-value angles by mixing a set of clusters with their respective predicted probabilities. Conversely, such good performance demonstrated that the distribution of protein backbone dihedral angles can be described through a set of clusters. Last but not the least, the everlasting development of deep learning models and optimization methods proves to be a powerful tool to promote new ideas and exploit new methods.

But there is still room for improvement. RaptorX-Angle just used one-dimensional features and adopted 1D CNN. It cannot extract information of long range interaction. Heffernan \emph{et al.} has developed more accurate SPIDER3 employing Long Short-Term Memory (LSTM) Bidirectional Recurrent Neural Networks (BRNNs), which are capable of capturing long range interactions \cite{heffernan2017capturing}. That is, considering pairwise interaction can further increase prediction accuracy. We will include two-dimensional features and exploit 2D CNN to see how much improvement could be achieved.

Moreover, as mentioned before, accumulation of prediction errors has buried the usefulness of torsion angles to construct 3D models. There is a great demand to develop a proper technique to deal with the errors. A general pipeline to add angle restraints and confidence to improve protein tertiary structure prediction need to be developed.

In conclusion, this study has made a more reliable prediction of dihedral angles and may facilitate protein structure prediction and functional study. In the near future, we can use the angle restraints to do tertiary structure prediction, which should be considered carefully to deal with errors and flexibility. We can also adopt the angle prediction to aid structure alignment and fold recognition.

\section*{Acknowledgements}
This work has been partly supported by the National Key Research and Development Program of China(No.2016YFA0502303), the National Key Basic Research Project of China (No. 2015CB910303), the National Natural Science Foundation of China (No.31471246) and China Scholarship Council.
It is also supported by National Institutes of Health grant R01GM089753 to JX and National Science Foundation grant DBI-1564955 to JX. The authors are also grateful to the support of Nvidia Inc.


\bibliography{AnglePrediction}

\end{document}